\documentclass[prd,twocolumn,showpacs,amsmath,amssymb,superscriptaddress,preprintnumbers,nofootinbib,showkeys]{revtex4}

\begin{document}

\newcommand{\nablab}{{\mathop {\rule{0pt}{0pt}{\nabla}}\limits^{\bot}}\rule{0pt}{0pt}}

\title{Towards nonlinear axion-dilaton electrodynamics: \\ How can axionic dark matter mimic dilaton-photon interactions?}

\author{Alexander B. Balakin}
\email{Alexander.Balakin@kpfu.ru} \affiliation{Department of
General Relativity and Gravitation, Institute of Physics, Kazan
Federal University, Kremlevskaya str. 16a, Kazan 420008, Russia}

\author{Aliya A. Galimova}
\email{aliya.galimova@kpfu.ru} \affiliation{Department of
General Relativity and Gravitation, Institute of Physics, Kazan
Federal University, Kremlevskaya str. 16a, Kazan 420008, Russia}

\date{\today}

\begin{abstract}
In the framework of the Einstein-Maxwell-axion-aether theory we establish the model, the Lagrangian of which contains the sin-type generalization of the term describing the axion-photon coupling, and the axionically induced cosine-type modification of the term attributed to the dilaton-photon interactions. The extension of the axion-dilaton-aether electrodynamics is inspired by the Jackson's idea concerning the internal symmetry of the equations of electromagnetism.
The application of the extended theory to the anisotropic homogeneous cosmological model  of the Bianchi-I type is considered. The exact solutions to the model evolutionary equations are obtained for the case, when the axionic dark matter is in the state of equilibrium, which is characterized by vanishing potential of the pseudoscalar field and its first derivative. The state of the axion-photon system, which  is of a new type and is indicated as a dynamic equilibrium, is studied in the framework of electrodynamics with axionic non-linearity. We show that the nonlinear axion-photon interactions can mimic the dilaton-photon coupling. We discuss the stability of the model with respect to homogeneous fluctuations of the axion field.

\end{abstract}
\pacs{04.20.-q, 04.40.-b, 04.40.Nr, 04.50.Kd}
\keywords{Alternative theories of gravity, axion electrodynamics, dynamic aether}
\maketitle

\section{Introduction}

\subsection{On the structure of the nonlinear axion-dilaton electrodynamics}

The formalism of nonlinear electrodynamics  is well elaborated and documented \cite{1,2}; it has a lot of applications to the models, which describe the physical systems with strong electromagnetic field (see, e.g., \cite{3}). The theory of gravity involves into consideration the sources, for the description of which the nonlinear electrodynamics is necessary, at least in three cases: when it deals with the astrophysical objects with strong electromagnetic fields \cite{5,6,7,8,9,10}), with early cosmology in the presence of  strong primordial magnetic fields \cite{11,12,121}), and when the propagation of nonlinear waves is under study \cite{P1,P2}. This formalism is based on the work with one true invariant of the electromagnetic field, $I_1 {=} F_{mn}F^{mn}$, and one pseudo-invariant $I_2 {=} F^{*}_{mn}F^{mn}$. The term $F_{mn}$ denotes the Maxwell tensor, and $F^{*}_{mn}$ is its dual. Starting from the well-known Born-Infeld Lagrangian (see, e.g., \cite{1} for historical details) physicists have invented a lot of special nonlinear  Lagrangians (exponential, power-law, fractional, logarithmic, etc.), which can be generally formulated as ${\cal L}(I_1,I_2^2)$ (see, e.g., \cite{5,6,7,8,9,10,11,12,121}). The theory of interaction between the electromagnetic and scalar fields has involved into the Lagrangian the terms of the form $I_3=f(\varphi) F_{mn}F^{mn}$, where $\varphi$ describes the scalar field, and the function $f(\varphi)$ is the subject of mathematical modeling. The theories linear in the electromagnetic field and nonlinear in the scalar field, have been indicated as Maxwell-dilaton theories (see, e.g., \cite{13,14,15,16,17,18,19,20}). The new true scalar $I_4=\phi F^{*}_{mn}F^{mn}$, where $\phi$ is the pseudoscalar field, appeared in the electromagnetic theory in the work \cite{WTNi77}. Due to the prediction of new light massive pseudo-Goldstone bosons, indicated later as axions, \cite{PQ,Weinberg,Wilczek}, and due to rapid development of this theory  we have now a rather branched axion electrodynamics (see. e.g., \cite{a1,a2,a3,a4,a5,a6,a7,a8}). There exists now the tendency for the unification of the Maxwell-dilaton theory and axion electrodynamics \cite{U1,U2,U3}), and for formulation of the nonlinear version of this unified theory \cite{NU1,NU2}. In other words, the relativistic astrophysics and cosmology, which operate with the axionic dark matter \cite{ADM1,ADM2,ADM3} engage the theoreticians for construction of the nonlinear axion-dilaton electrodynamics. If this theory is planned to be formulated on the base of standard Lagrange formalism, it could be equipped by four invariants, listed above: $I_1$, $I_2$, $I_3$, $I_4$, and/or various combinations of these quantities. However, if we act in the spirit of Occam's razor concept, we can formulate the Lagrangian of the nonlinear axion-dilaton electrodynamics ${\cal L}_{(\rm U)}$ based on appropriate function of one generalized (unified) invariant ${\cal I}$, say as ${\cal L}_{(\rm U)}({\cal I})$. We are ready to postulate this Lagrangian, however, in order to motivate our choice, we invite readers to recall the idea of internal (latent) invariance of classical linear electrodynamics, advocated by  Jackson in the textbook \cite{Jackson}.

\subsection{Prologue: On the internal (latent) invariance of the classical linear electrodynamics}

Equations of vacuum electrodynamics are known to be symmetric with respect to linear transformation of electric and magnetic fields \cite{Jackson}. This fact can be illustrated, for instance, as follows. Let the vacuum Maxwell equations be written using the Maxwell tensor $F^{ik}$ and its  dual $F^{*ik}$ in the form
\begin{equation}
\nabla_k F^{ik}=0 \,, \quad \nabla_k F^{*ik}=0 \,,
\label{1}
\end{equation}
where $F^{*ik} {=} \frac12 \epsilon^{ikmn} F_{mn}$, with the Levi-Civita pseudotensor $\epsilon^{ikmn} {=} \frac{E^{ikmn}}{\sqrt{-g}}$, ($E^{0123}=1$). One can consider the linear combination of these two tensors
\begin{equation}
{\cal F}^{ik} = F^{ik} \cos{\left(\frac{\alpha}{2}\right)}  + F^{*ik} \sin{\left(\frac{\alpha}{2}\right)}
\label{2}
\end{equation}
with a constant angle value $\alpha$, and check that the equations
\begin{equation}
\nabla_k {\cal F}^{ik}=0 \,, \quad \nabla_k {\cal F}^{*ik}=0
\label{3}
\end{equation}
are valid due to the equations (\ref{1}).
Let us consider the electric field four-vector ${\cal E}^i$ and the four-vector of the magnetic induction ${\cal B}^i $, given by the standard relationships
\begin{equation}
{\cal E}^i \equiv {\cal F}^{ik}U_k \,,  \quad {\cal B}^i \equiv {\cal F}^{*ik}U_k \,,
\label{012}
\end{equation}
where $U^k$ is the velocity four-vector, associated with some observer.
Then one can obtain from (\ref{2}) the following relationships:
$$
{\cal E}^i =  E^{i} \cos{\left(\frac{\alpha}{2}\right)} +  B^{i} \sin{\left(\frac{\alpha}{2}\right)} \,,
$$
\begin{equation}
{\cal B}^i =  - E^{i} \sin{\left(\frac{\alpha}{2}\right)}  + B^{i} \cos{\left(\frac{\alpha}{2}\right)} \,,
\label{4}
\end{equation}
which describe, in fact, the rotation in the plane  $E \times B$ with respect to the angle $\frac12 \alpha$.
The Lagrangian constructed using the term ${\cal F}^{ik}$
\begin{equation}
{\cal L}= \frac14 {\cal F}^{ik}{\cal F}_{ik} \,,
\label{5}
\end{equation}
can be rewritten in the following form:
\begin{equation}
{\cal L}= \frac14 \left( F_{ik}F^{ik} \ \cos{\alpha} +  F^{ik}F^*_{ik} \ \sin{\alpha} \right) \,.
\label{6}
\end{equation}
In other words, we obtain the Lagrangian which admits internal discrete symmetry under the transformation $\alpha \to \alpha + 2\pi n $ ($n$ is an integer).
The authors of the axion electrodynamics have introduced the pseudoscalar field $\theta(x)$ depending on coordinates, but they also assumed that the theory is invariant with respect to the discrete symmetry $\theta(x) \to \theta(x) {+} 2\pi n$. In this context it is logical to replace $\alpha$ in (\ref{6}) with $\theta(x)$ and consider the properties of such Lagrangian.

First of all, when $\theta \to 0$, we recover the structure of the classical Lagrangian of the axion electrodynamics
\begin{equation}
{\cal L} \to  \frac14 \left[F_{ik}F^{ik} + \theta F^{ik}F^*_{ik}\right] \,,
\label{8}
\end{equation}
i.e., we deal with the specific sin-type nonlinear generalization of the Lagrangian of the axion field.

Second, $\sin{\theta(x)}$ is the odd function, and thus the term $\frac14 F^{ik}F^*_{ik} \cdot \sin{\theta}$ also is the true invariant.

Third, $\cos{\theta(x)}$ is the even function, thus the first term in (\ref{6}) can be represented as $\frac14 F_{ik}F^{ik} \cdot \cos{|\theta(x)|}$. If we imagine that $|\theta|=\varphi$, i.e., it can play the role of a scalar, this expression looks like the specific (cosin-type) Lagrangian describing the dilaton-photon interactions, with $f(\varphi)= \cos{\varphi}$.

Finally, we can suggest that the unified Lagrangian
$$
{\cal L}_{(\rm U)} = {\cal L}_{(\rm U)}({\cal I}) \,,
$$
\begin{equation}
{\cal I} \equiv \frac14 \left[F_{ik}F^{ik} \cos{\theta(x)} {+}  F^{ik}F^*_{ik} \sin{\theta(x)} \right]
\label{qqq}
\end{equation}
can be chosen as one of the versions appropriate for the  nonlinear axion-dilaton electrodynamics. We have to emphasize, that based on the internal symmetry of classical electrodynamics we can motivate the description of both: the dilaton and axion contributions, using only one pseudoscalar field.

\subsection{Motivation of the work}

Models of the Universe in the modern theoretical cosmology include necessarily  two dark substrata in addition to the standard elements: gravitational, electromagnetic fields and ordinary matter. The first dark substratum, the dark matter, appeared as the element explaining the distribution of the rotation velocities in the spiral galaxies. The second substratum, the dark energy, was introduced to explain the late-time accelerated expansion of the Universe. These two basic dark substrata have to interact with the gravitational, electromagnetic fields, with the ordinary matter and with each other. Probably the direct coupling between the dark substrata exists, and the description of the mechanism of this coupling requires to identify the nature of them. Many scientists admit that the dark matter consists of axions, and the properties of the axion - photon coupling are well-documented; in our work we follow this idea. The nature of the dark energy causes a lot of discussions; from the series of models we prefer the model of dynamic aether \cite{AE1,AE2,AE3,AE4,AE5,AE6}, which could explain the accelerated expansion of the Universe and gives additional bonuses in the Universe expansion modeling (this dark element provides the theory to have the intrinsic preferred velocity field \cite{Pref1,Pref2}, though it translates the theory to the category of the ones with Lorentz violation \cite{LV1,LV2,LV3}).
In this sense, when the unit vector field $U^i$ associated with the aether velocity, and the pseudoscalar field $\phi$ associated with the axionic dark matter appear
in the Lagrangian of our theory, it is motivated by the completeness of the cosmological model.

Since we admit that the dark matter interacts with the dark energy, we propose to study a new mechanism of this coupling. To be more precise, instead of a featureless standard model of two-component dark fluid, we consider the very clear model, in which the expansion scalar $\Theta = \nabla_kU^k$ of the aether flow is installed into the potential of the pseudoscalar (axion) field $V(\phi, \Theta)$. From the mathematical point of view, this installation can be made using a new guiding function depending on $\Theta$; we indicate it below as $\Phi_*(\Theta)$ and write the potential as follows $V(\phi, \Phi_*) = V_0\left[1-\cos{\left(\frac{2\pi \phi}{\Phi_*} \right)} \right]$. We have to stress, that the guiding function $\Phi_*$ is not a new dynamic variable; the variation of the Lagrangian with respect to $\Phi_*$ is not provided, and thus there is no additional equation for the evolution of this function.

What is the search method for the quantity $\Phi_*$? We use the analogy with the  kinetic theory of the relativistic gas, based on the kinetic equation \cite{KT}
\begin{equation}
p^i \frac{\partial f}{\partial x^i} - \Gamma^i_{kl} \frac{p^k p^l}{m} \frac{\partial f}{\partial p^i} = I(f) \,,
\label{kineq1}
\end{equation}
which includes the distribution function $f(x,p^k)$ depending on coordinates and momentum four-vector $p^k$, the Christoffel symbols $\Gamma^i_{kl}$ and the collision integral $I$.
There are four details, which allow us to speak about similarity.

\noindent
a) The equilibrium distribution function $f_{(\rm eq)}(x,p^k)$ can be found from the condition that the collision integral vanishes $I(f_{(\rm eq)})=0$. In the theory of the axion evolution we search for the equilibrium value of the pseudoscalar field $\phi_{(\rm eq)}$ from the condition that the potential itself and its derivative vanish, $V(\phi_{(\rm eq)})=0 $ and $\frac{\partial V}{\partial \phi}(\phi_{(\rm eq)})=0$.

\noindent
b) In the kinetic theory this procedure defines the equilibrium distribution function as an arbitrary function of the chemical potential $\mu(x)$, of the temperature $T(x)$ and of the unit four-vector $V^i$ attributed to the macroscopic velocity of the system as a whole. In the classical formula
\begin{equation}
f_{(\rm eq)}(x,p^k) = \frac{h^{-3} \rho}{e^{\cal U}\pm 1} \,, \quad {\cal U} = \frac{-\mu + p^k V_k}{k_{B} T}
\label{kineq2}
\end{equation}
$h$ is the Planck constant, $\rho$ is the degeneracy factor, and the scalar function ${\cal U}$ contains the Boltzmann constant $k_B$. The sign plus relates to the fermion distribution, and the sign minus corresponds to bosons. The construction of the equilibrium axion function $\phi_{(\rm eq)}= n \Phi_*$ ($n$ is an integer) is much more simple: it contains only one arbitrary function, namely, the guiding function $\Phi_*(x)$.

\noindent
c) Then in the kinetic theory it is necessary to put the equilibrium function (\ref{kineq2}) into the left-hand side of the kinetic equation (\ref{kineq1}) to obtain the unknown functions $\mu(x)$, $T(x)$ and $V^i(x)$ from the equations of the first and second moments evolution (the particle number density four-vector and the stress-energy tensor). In the case with the axion field we put the equilibrium function $\phi_{(\rm eq)} = n \Phi_*$, first, into the evolutionary equation for the pseudoscalar field, second, into the equations of axion electrodynamics, third, into the equations for the gravity field. In other words, we obtain the guiding function $\Phi_*$ from the set of self-consistent equations in the assumption that the axion field is in the equilibrium state.

\noindent
d) The next step in the kinetic theory is to study the states near the equilibrium; for this purpose one considers the fluctuations of the distribution function $f= f_{(\rm eq)} + \delta f$ and one studies the solution to the linearized kinetic equation for $\delta f$. In the axion theory we consider $\phi = n \Phi_* + \psi$, and analyze the evolution of the fluctuation of the pseudoscalar field $\psi$.

To conclude, we have to emphasize, that to search for the guiding function $\Phi_*$ we use not a specific trick but the standard receipt, which is well-tested in the kinetic theory. As for the problem of initial data, usually, in the kinetic theory the equilibrium chemical potential and temperature are fixed on some spacetime hyper-surface based on some physical requirements; in the axion theory, the initial data for the guiding function are also fixed and can be the subject of physical modeling.

It is important to stress that in the presented work the Lagrangian is suitably prepared; this means that the potential $V(\phi, \Phi_*(\Theta))$ as well as the multipliers $\cos{\left(\frac{2\pi \phi}{\Phi_*}\right)}$ and $\sin{\left(\frac{2\pi \phi}{\Phi_*}\right)}$, which depend on the expansion scalar $\Theta= \nabla_k U^k$, are ready to participate in the variation of the action functional with respect to the vector field $U^i$ and with respect to the metric $g_{ik}$. As the result, new terms appear in the evolutionary equations for the aether velocity and in the equations for the gravity field. In this sense, the search for the guiding function is the well-defined self-consistent procedure.

Since the axionic field interacts with the electromagnetic one, we proposed to study some new model of nonlinear axion-photon coupling. In this model the aether vector field also interacts with the electromagnetic one, but now it is an indirect type of coupling, it is the so-called aether guidance of the axion-photon interaction.

Keeping in mind these modifications of the theory, we can answer three natural questions of the readers: were
these modifications necessary to solve some problem, is this model meaningful and what is the merit of this study?
Let us imagine the anisotropic homogeneous early Universe with strong magnetic field \cite{MAG}. Interaction of the magnetic field with the non-stationary axionic dark matter inevitably produces the  electric field parallel to the magnetic one. The axionicaly induced electric field accelerates the charged particles in the direction pointed by the magnetic field; when the magnetic field is strong, the electric field also can be rather intensive, and the accelerated particle can become ultrarelativistic. The accelerated charges emit, and, as the theory of the electromagnetic radiation predicts, for the ultrarelativistic particles the radiation concentrates in small angle near the propagation direction.
Thus the anisotropy of the photon distribution appears, which marks the directions of primordial magnetic fields. In other words, one can assume that the well-known portrait of the temperature fluctuations, presented by the WMAP mission, depicts not only the standard results of normally distributed matter fluctuations, but displays also some initial distribution of the magnetic fields in the early Universe. Of course, this idea requires the detailed data analysis, however, it could permit to take an alternative look at the problem.

But the main merit of the model, constructed in our work, is the following. When the primordial magnetic field is strong, one has to work with the nonlinear electrodynamics. We considered the simple model of the nonlinear axion electrodynamics and have found that the mentioned axionically induced electric field can, in principle, abnormally grow in accordance with the law $\propto \tan{\left(\frac{2\pi \phi}{\Phi_*}\right)}$. Clearly, if the pseudoscalar field $\phi$ reaches the critical value $\phi = \frac14 \Phi_*$, the electric field, the invariants of the electromagnetic field, etc. take infinite values.
In other words, the nonlinear Einstein-Maxwell-axion model is potentially instable. The evident advantage of our model is that the found guiding function $\Phi_*$ removes this catastrophic scenario, or equivalently, the wise guidance of the aether prevents the uncontrolled anisotropisation of the Universe. One can add that even in the framework of the relativistic kinetic theory based on the linear axion electrodynamics there exist instable submodels (see, e.g., \cite{Vlasov}); the nonlinear axion electrodynamics seems to be rich in examples of instabilities.

\subsection{How is the work organized?}

Before we establish the full-format nonlinear theory (we hope to do it in the next work), we would like to check the results of the proposed unified approach using the truncated model, in which the pseudoscalar (axion) field is nonlinear (of the sin-type), the induced dilaton field is also nonlinear (of the cosin-type), however the electromagnetic field is described by the linear equations.
The formalism of the dynamic aether is presented in the linear version also.

The paper is organized as follows. In Section II  we describe the formalism of the extended Einstein-Maxwell-aether-axion theory: we introduce the extended Lagrangian, the modified periodic potential of the axion field, we derive the master equations for the electromagnetic, gravitational, pseudoscalar and unit vector fields. In Section III we consider the cosmological application of the theory using the anisotropic homogeneous Bianchi-I model as an example. Section IV contains discussion and conclusions.

\section{Modified Einstein-Maxwell-aether-axion model}

The axion electrodynamics works with the dimensionless pseudoscalar (axion) field $\phi$, which is linked with the quantity  $\theta(x)$ by the relationship $\theta = \frac{2\pi \phi}{\Phi_*}$. We assume that both: the pseudoscalars $\phi$ and $\Phi_*$ are dimensionless, and $\Phi_*$ describes a vacuum averaged value of the axion field. This idea was motivated and used, e.g., in \cite{Equilib1,Equilib2,Equilib3,Equilib4,Equilib5}). Based on this idea we consider now  the following unified invariant ${\cal I}_{(U)}$ nonlinear in the pseudoscalar field:
\begin{equation}
 \frac14 \left[F_{ik}F^{ik} \cos{\left(\frac{2\pi \phi}{\Phi_*}\right)} {+} \left(\frac{\Phi_*} {2\pi}\right)  F^{ik}F^*_{ik} \sin{\left(\frac{2\pi \phi}{\Phi_*}\right)}\right].
\label{7}
\end{equation}
The multiplier in front of the second term is necessary, since when the pseudoscalar field is weak, i.e., $\left(\frac{2\pi \phi}{\Phi_*}\right)\to 0$ just the invariant (\ref{7}) takes the form
\begin{equation}
{\cal I}_{(U)} \to  \frac14 \left[F_{ik}F^{ik} + \phi F^{ik}F^*_{ik}\right] \,,
\label{8}
\end{equation}
advocated by Wei-Tou Ni in \cite{WTNi77}.

\subsection{The total action functional}

We consider the model of interaction of four fields; the gravitational and unit vector fields are the key elements of the Einstein-aether theory \cite{AE1,AE2,AE3,AE4,AE5,AE6}; the electromagnetic and the pseudoscalar fields appear as the key elements of the axion electrodynamics \cite{PQ} - \cite{a5}. Respectively, we divide the total action functional into two parts:
\begin{equation}
S_{(\rm total)} = S_{(\rm EA)} +  S_{(\rm MA)} \,.
\label{9}
\end{equation}
The first term given by
$$
-S_{(\rm EA)} = \int d^4x  \frac{\sqrt{-g}}{2\kappa}\left[R{+}2\Lambda + \lambda\left(g_{mn}U^mU^n - 1 \right) +  \right.
$$
\begin{equation}
\left. + K^{ab}_{\ \ mn} \nabla_a U^m  \nabla_b U^n \right] \,,
\label{10}
\end{equation}
is the standard action functional of the Einstein-aether theory, it contains
the determinant of the metric $g$, the covariant derivative $\nabla_k$, the Ricci scalar $R$, the cosmological constant $\Lambda$, the Einstein constant  $\kappa$, the Lagrange multiplier $\lambda$, the unit timelike vector field $U^i$, which is associated with the velocity four-vector of the aether flow.
The constitutive tensor
\begin{equation}
K^{ab}_{\ \ mn} = C_1 g^{ab}g_{mn} {+} C_2 \delta^a_m \delta^b_n {+} C_3 \delta^a_n \delta^b_m {+} C_4 U^a U^b g_{mn}
\label{11}
\end{equation}
contains four phenomenological constants $C_1$, $C_2$, $C_3$, $C_4$.
The second term in (\ref{9}) can be represented as follows:
\begin{equation}
-S_{(\rm MA)} = \int d^4x \sqrt{-g} \left\{\frac12 \Psi^2_0   \left[V {-} \nabla_k \phi \nabla^k \phi \right]  + \right.
\label{11}
\end{equation}
$$
\left. {+} \frac14 \left[\cos{\left(\frac{2\pi \phi}{\Phi_*}\right)} F_{ik}F^{ik} {+} \left(\frac{\Phi_*} {2\pi}\right) \sin{\left(\frac{2\pi \phi}{\Phi_*}\right)} F^{ik}F^*_{ik}\right] \right\} \,.
$$
It describes the electromagnetic field of the Maxwell type interacting with the nonlinear pseudoscalar (axion) field.
Here $V$ is the potential of the pseudoscalar (axion) field; the parameter $\Psi_0$ relates to the coupling constant of the axion-photon interaction $g_{A \gamma \gamma}$, $\frac{1}{\Psi_0}=g_{A \gamma \gamma}$.

\subsection{The modified periodic potential of the axion field}

The potential of the axion field inherits the discrete symmetry, which in our terms can be indicated as $\frac{2\pi \phi}{\Phi_*} \to \frac{2\pi \phi}{\Phi_*} + 2\pi n$; therefore,
we can use the modified periodic potential in the form (see, e.g., \cite{Equilib1,Equilib2,Equilib3,Equilib4,Equilib5}):
\begin{equation}
V(\phi,\Phi_{*}) = \frac{m^2_A \Phi^2_{*}}{2\pi^2} \left[1- \cos{\left(\frac{2 \pi \phi}{\Phi_{*}}\right)} \right] \,.
\label{13}
\end{equation}
This periodic potential has the minima at $\phi=n \Phi_{*}$. Near the minima, when
$\phi \to n \Phi_{*} {+} \psi$ and $|\psi|$ is small, the potential takes the standard form $V \to m^2_A \psi^2$, where $m_A$ is the axion rest mass.
The guiding function $\Phi_{*}$ is assumed to depend on the scalar $\Theta$, i.e., $\Phi_{*} = \Phi_{*}(\Theta)$, where $\Theta \equiv \nabla_k U^k$ is the expansion scalar of the aether flow. In this sense, below we use the notation $V(\phi,\Theta)$.

\subsection{Master equations of the model}

\subsubsection{Master equations for the unit vector field}

Standard variations with respect to the Lagrange multiplier $\lambda$ and to the four-vector $U^i$ give the following set of equations
\begin{equation}
g_{mn}U^m U^n =1 \,,
\label{14}
\end{equation}
\begin{equation}
\nabla_a {\cal J}^{a}_{\ j}  = \lambda  U_j  +  C_4 DU_m \nabla_j U^m -
\kappa \nabla_j \left(\frac{d \Phi_*}{d \Theta} \Omega \right) \,.
\label{15}
\end{equation}
Here the tensor ${\cal J}^{a}_{ \ j}$ is standardly presented as
$$
{\cal J}^{a}_{ \ j} = K^{ab}_{\ \ jn} \nabla_b U^n =
$$
\begin{equation}
= C_1 \nabla^a U_j + C_2 \delta^a_j \Theta + C_3\nabla_j U^a + C_4 U^a DU_j  \,,
\label{16}
\end{equation}
with the convective derivative $D = U^k \nabla_k$.
The scalar $\Omega$ can be written as follows:
$$
\Omega = - \frac{\phi}{4\Phi_*}F^*_{pq}F^{pq} \cos{\left(\frac{2 \pi \phi}{\Phi_{*}}\right)}
+
$$
$$
+\left[\frac{\pi \phi}{2\Phi^2_*} F_{pq}F^{pq} + \frac{1}{8\pi}F^*_{pq}F^{pq} \right] \sin{\left(\frac{2 \pi \phi}{\Phi_{*}}\right)} +
$$
\begin{equation}
+\frac{m^2_{A}\Psi^2_0 }{2\pi^2}  \left\{ \Phi_{*}\left[1{-} \cos{\left(\frac{2 \pi \phi}{\Phi_{*}}\right)} \right]
{-}\pi \phi  \sin{\left(\frac{2 \pi \phi}{\Phi_{*}}\right)} \right\}.
\label{17}
\end{equation}

\subsubsection{Master equations for the electromagnetic field}

Variation of the total action functional with respect to the electromagnetic potential $A_i$ gives the equations
\begin{equation}
\nabla_k H^{ik} = 0 \,,
\label{18}
\end{equation}
where the excitation tensor $H^{ik}$ is now of the form
\begin{equation}
H^{ik} = \cos{\left(\frac{2 \pi \phi}{\Phi_{*}}\right)} F^{ik} + \frac{\Phi_{*}}{2 \pi} \sin{\left(\frac{2 \pi \phi}{\Phi_{*}}\right)} \ F^{*ik}  \,.
\label{19}
\end{equation}
Taking into account the equation
\begin{equation}
\nabla_k F^{*ik}=0 \,,
\label{20}
\end{equation}
which converts into identity, if one uses the definition of the Maxwell tensor
\begin{equation}
F_{ik} = \nabla_{i}A_k - \nabla_k A_i \,,
\label{21}
\end{equation}
one can rewrite the equation (\ref{18}) as follows
\begin{equation}
 \nabla_k \left[\cos{\left(\frac{2 \pi \phi}{\Phi_{*}}\right)} F^{ik}\right] {=} {-} F^{*ik} \nabla_k \left[\frac{\Phi_{*}}{2 \pi} \sin{\left(\frac{2 \pi \phi}{\Phi_{*}}\right)} \right].
\label{22}
\end{equation}
When the axion field is weak, (\ref{22}) converts into the standard equation of axion electrodynamics
\begin{equation}
 \nabla_k F^{ik}= - F^{*ik} \nabla_k  \phi \,,
\label{23}
\end{equation}
derived in \cite{WTNi77}.

\subsubsection{Master equation for the axion field}

Variation of the total action functional with respect to the axion field yields
\begin{equation}
\nabla^k \nabla_k \phi {+} \frac{m^2_A \Phi_{*}}{2\pi} \sin{\left(\frac{2 \pi \phi}{\Phi_{*}}\right)}  =
\label{24}
\end{equation}
$$
{=} \frac{1}{4\Psi^2_0} \left[\frac{2 \pi}{\Phi_{*}} \sin{\left(\frac{2 \pi \phi}{\Phi_{*}}\right)}  F_{ik}F^{ik} {-} \cos{\left(\frac{2 \pi \phi}{\Phi_{*}}\right)}   F_{ik}F^{*ik} \right].
$$
There are two source-terms in the right-hand side of the equation (\ref{24}); for the model of weak axion field the second term gives the expression
$-\frac{1}{4\Psi^2_0} F_{ik}F^{*ik}$, which appears in the linear axion electrodynamics and is proportional to the second (pseudo)invariant of the electromagnetic field. The first term  represents in this approximation the source of a new type
$\frac{\pi^2 \phi}{\Psi^2_0 \Phi^2_{*}}  F_{ik}F^{ik}$, which is proportional to the first invariant of the electromagnetic field, and is linear in the axion field $\phi$.

\subsubsection{Master equations for the gravitational field}

Variation with respect to the metric gives the gravity field equations; they can be written in the form
\begin{equation}
R_{ik} - \frac12 R g_{ik} - \Lambda g_{ik} = T^{(\rm U)}_{ik} + \kappa T^{(\rm EMA)}_{ik} + \kappa T^{(\rm A)}_{ik} \,.
\label{25}
\end{equation}
The first term in the right-hand side of (\ref{25}) is the standard stress-energy tensor associated with the aether flow:
\begin{equation}
T^{(\rm U)}_{ik} =
\frac12 g_{ik} \ K^{abmn} \nabla_a U_m \nabla_b U_n{+} \lambda U_iU_k  {+}
\label{26}
\end{equation}
$$
{+}\nabla^m \left[U_{(i}{\cal J}_{k)m} {-}
{\cal J}_{m(i}U_{k)} {-}
{\cal J}_{(ik)} U_m\right]+
$$
$$
+C_1\left[(\nabla_mU_i)(\nabla^m U_k) {-}
(\nabla_i U_m )(\nabla_k U^m) \right]
{+}C_4 D U_i D U_k  \,.
$$
The second term does not contain functions attributed to the electromagnetic field
$$
T^{(\rm A)}_{ik}= \Psi^2_0 \left[\nabla_i \phi \nabla_k \phi  + \frac12 g_{ik}\left(V - \nabla_p \phi \nabla^p \phi \right) \right] +
$$
$$
+ \frac{m^2_A \Psi^2_0}{2\pi^2} g_{ik} \nabla_s \left\{U^s  \frac{d \Phi_{*}}{d\Theta}\left[\pi \phi \sin{\left(\frac{2 \pi \phi}{\Phi_{*}}\right)}  - \right. \right.
$$
\begin{equation}
 \left. \left. - \Phi_{*}\left[ 1-\cos{\left(\frac{2 \pi \phi}{\Phi_{*}}\right)}\right] \right]\right\} \,,
\label{27}
\end{equation}
it describes the contribution of the axion field coupled to the aether velocity via the expansion scalar $\Theta$, the argument of the axion potential (\ref{13}). The third term
$$
T^{(\rm MA)}_{ik} = \cos{\left(\frac{2\pi \phi}{\Phi_*} \right)}\left[\frac14 g_{ik} F_{pq}F^{pq} - F_{ip}F_k^{\ p}\right] +
$$
$$
+ \frac18 g_{ik} \nabla_s \left\{U^s  \frac{d \Phi_{*}}{d\Theta} \left[ \frac{\phi}{\Phi_*}F^{*}_{pq}F^{pq} \cos{\left(\frac{2 \pi \phi}{\Phi_{*}}\right)} -   \right. \right.
$$
\begin{equation}
\left. \left.  - \left(\frac{2\pi \phi}{\Phi^2_*} F_{pq}F^{pq} + \frac{1}{2\pi} F^{*}_{pq}F^{pq}\right) \sin{\left(\frac{2 \pi \phi}{\Phi_{*}}\right)} \right] \right\}
\label{28}
\end{equation}
contains both electromagnetic and axion fields coupled to the aether velocity via the expansion scalar $\Theta$, appeared in the course of nonlinear generalization of the axion electrodynamics.

\section{Cosmological application:  The Bianchi-I model}

The spatially homogeneous anisotropic cosmological model of the Bianchi-I type is characterized by the metric
\begin{equation}
ds^2 = dt^2 - a^2(t)dx^2 - b^2(t)dy^2 -c^2(t)dz^2 \,.
\label{29}
\end{equation}
We assume that all the physical quantities, which  describe the model, inherit the symmetry of this spacetime, in particular, they depend on the cosmological time $t$ only. Also, based on the symmetry of the model we consider the following assumptions: first, the aether velocity four-vector is of the form $U^i=\delta^i_0$; second, the magnetic and electric fields are parallel and are directed along the axis $0z$.
For this model the covariant derivative of the aether velocity four-vector is symmetric:
\begin{equation}
\nabla_iU_k = \nabla_kU_i = \frac12 \dot{g}_{ik} \,.
\label{30}
\end{equation}
Here and below the dot denotes the derivative with respect to cosmological time. As a consequence of (\ref{30}), the acceleration four-vector is vanishing, $DU_i=0$, and the expansion scalar is of the form
\begin{equation}
\Theta = \nabla_k U^k = \frac{\dot{a}}{a}+ \frac{\dot{b}}{b} + \frac{\dot{c}}{c} \,.
\label{31}
\end{equation}
The last detail of the model restriction is connected with the fact established by the observation of the events GW170817 and  GRB170817A \cite{GRB17}. The velocity of the gravitational waves is estimated to differ from the speed of light in vacuum by the quantity of the order $10^{-15}$. In fact, this means that the coupling constants $C_1$ and $C_3$ of the Einstein-aether theory have to satisfy the inequality $-6 \times 10^{-15}<C_1+C_3< 1.4 \times 10^{-15}$. Keeping in mind this estimation, we assume below that $C_1+C_3=0$.
Now we consider the reduced master equations of the model, and we start with the equations for the aether velocity.

\subsection{Reduced Master equations}

\subsubsection{Solutions to the Master equations for the aether velocity}

The tensor $J^{aj}$ defined by (\ref{16}) is now of the form
\begin{equation}
J^{aj} = C_2 \Theta g^{aj} \,,
\label{32}
\end{equation}
and thus, the reduced equations (\ref{15}) can be rewritten as follows:
\begin{equation}
\lambda U_j = \nabla_j \left(C_2 \Theta + \kappa \Omega \frac{d \Phi_*}{d \Theta} \right) \,.
\label{33}
\end{equation}
Three equations from the four presented by (\ref{33}) are satisfied identically, and the fourth (nontrivial) equation gives the Lagrange multiplier
\begin{equation}
\lambda(t) = C_2 \dot{\Theta}
+ \frac{\kappa}{8\pi} \frac{d}{dt}\left\{\frac{d\Phi_*}{d\Theta}\left[-Y \cos{Y} F^*_{pq}F^{pq}   + \right. \right.
\label{34}
\end{equation}
$$
\left. \left. + \sin{Y}
\left[\left(\frac{2\pi}{\Phi^*}\right) Y F_{pq}F^{pq} + F^*_{pq}F^{pq}  \right] + \right. \right.
$$
$$
\left. \left. + 8 m^2_A \Psi^2_0 \left(\frac{\Phi_*}{2\pi}\right) \left(1-\cos{Y} - \frac12 Y \sin{Y} \right)\right] \right\}  \,.
$$
Here we introduced the auxiliary function
\begin{equation}
Y(t) =  \frac{2\pi\phi}{\Phi_*} \,,
\label{35}
\end{equation}
which is used below for convenience.

\subsubsection{Solutions to the electrodynamic equations}

When the potential of the electromagnetic field depends on the cosmological time only, the equations (\ref{20}) give the solution
\begin{equation}
F_{12}=const = B_0 \,.
\label{36}
\end{equation}
Then we reduce the equations (\ref{22})  to
\begin{equation}
\frac{d}{dt} \left\{abc\left[F^{i0} \cos{Y}+ \left(\frac{\Phi_*}{2\pi}\right) F^{*i0} \sin{Y} \right] \right\} =0 \,,
\label{37}
\end{equation}
the solution to which is
\begin{equation}
F^{30} \cos{Y}- \frac{\Phi_*}{2\pi abc} F_{12} \sin{Y} = \frac{{\rm const}}{abc} \,.
\label{38}
\end{equation}
The quantity $E(t) \equiv c(t) F^{30}$ plays the role of the electric field; keeping in mind the initial data at $t=t_0$, we obtain from (\ref{38}) that
\begin{equation}
E(t) = E(t_0) \frac{a(t_0) b(t_0) \cos{Y(t_0)}}{a(t) b(t) \cos{Y(t)}} +
\label{39}
\end{equation}
$$
+ \frac{B_0}{2\pi a(t)b(t)\cos{Y(t)}} \left[\Phi_*(t) \sin{Y(t)} {-} \Phi_*(t_0) \sin{Y(t_0)}\right] \,.
$$
Below we consider the case, when the initial electric field in the Universe was absent, $E(t_0)=0$, and the global magnetic field was the only source of the Universe anisotropy.
Also we assume that at the same time moment, $t=t_0$, the quantity $Y(t_0)$ is proportional to $2\pi$, i.e., $Y(t_0)= 2\pi n_0$, where $n_0$ is some integer. Thus the formula (\ref{39}) takes the following elegant form:
\begin{equation}
E(t) =   \frac{B_0 \Phi_*(t)}{2\pi a(t)b(t)} \tan{Y(t)} \,.
\label{40}
\end{equation}
For the electromagnetic field with this structure we obtain two invariants of the electromagnetic field:
\begin{equation}
F^*_{pq}F^{pq} = \frac{4B_0^2}{a^2 b^2} \left(\frac{\Phi_*}{2\pi}\right) \tan{Y} \,,
\label{41}
\end{equation}
\begin{equation}
F_{pq}F^{pq} = \frac{2B_0^2}{a^2 b^2}\left[1-  \left(\frac{\Phi_*}{2\pi} \right)^2 \tan^2{Y}\right] \,.
\label{42}
\end{equation}
Based on the solutions to the equations of electrodynamics we can specify the equations for the axion field.

\subsubsection{Reduced equation for the axion field}

The axion field equation (\ref{24}) takes now the form
\begin{equation}
\ddot{\phi} + \Theta \dot{\phi} =
\label{43}
\end{equation}
$$
=  \left(\frac{\Phi_*}{2\pi}\right) \sin{Y} \left\{\frac{B^2_0}{2\Psi^2_0 a^2 b^2}\left[\left(\frac{2\pi}{\Phi_*}\right)^2 {-} 2 {-} \tan^{2}{Y} \right] {-} m^2_A \right\} \,.
$$
This equation is nonlinear, and it does not give a chance to solve (\ref{43}) analytically in general case. However, there are three special cases for which the right-hand side of this equation vanishes.

\noindent
1. {\it Axionic dark matter is in a standard equilibrium state.}

\noindent
Let us assume that $\phi(t) = n \Phi_*(t)$; in this case the axion potential (\ref{13}) and its derivative vanish, and thus we speak about the equilibrium state of the axionic system \cite{Equilib1}-\cite{Equilib5}. Now $Y=2 \pi n$, $\sin{Y}=0$ and thus the right-hand side of the equation (\ref{43}) takes zero value. Mention that the requirement $\phi(t) = n \Phi_*(t)$ itself does not restrict the choice of the guiding function $\Phi_*(t)$, however, the remaining equation $\ddot{\Phi_*} + \Theta \dot{\Phi_*} =0$ predetermines the behavior of $\Phi_*(t)$.

\noindent
2. {\it Axionic dark matter is in a dynamic equilibrium state.}

\noindent
Let us assume that $\frac{B^2_0}{2\Psi^2_0 a^2 b^2}\left[\left(\frac{2\pi}{\Phi_*}\right)^2 {-} 2 {-} \tan^{2}{Y} \right] {=} m^2_A $.
In this case the potential (\ref{13}) does not vanish, however, the contribution of the axion potential to the equation  (\ref{43}) is compensated by the contribution of the electromagnetic field.

\noindent
3. {\it Axionic dark matter is in a mixed equilibrium state.}

\noindent
Let us assume that the requirements, formulated in the items 1. and 2., are fulfilled simultaneously; then we obtain the guiding function in the explicit form
\begin{equation}
\Phi_* = \frac{\pi \sqrt2}{\sqrt{1+ \frac{\Psi^2_0 m^2_A a^2 b^2}{B^2_0}}} \,.
\label{143}
\end{equation}
Clearly, this function has to be constant and the corresponding spacetime should be quasi-static with $ab=const$. Below we will discuss all three cases in more detail.

\subsubsection{Reduced equations of the gravity field: Preliminaries}

Now we consider the stress-energy tensors (\ref{26}), (\ref{27}) and (\ref{28}), reduced with respect to the symmetry of the Bianchi-I model.
First of all we write the stress-energy tensor, which is attributed to the dynamic aether
$$
T^{(\rm U)}_{ik} = - C_2 \left[\frac12 \Theta^2 g_{ik} + \dot{\Theta}\left(g_{ik}-U_iU_k \right) \right] +
$$
$$
+ \frac{\kappa}{8\pi} U_i U_k  \frac{d}{dt}\left\{\frac{d\Phi_*}{d\Theta}\left[\left(\frac{2\pi}{\Phi_*}\right) Y \sin{Y} F_{pq}F^{pq} + \right. \right.
$$
$$
\left. \left. + F^{*}_{pq}F^{pq} \left(\sin{Y}-Y \cos{Y} \right) + \right. \right.
$$
\begin{equation}
\left. \left.
+8 m^2_A \Psi^2_0 \left(\frac{\Phi_*}{2\pi}\right) \left(1{-}\cos{Y} {-} \frac12 Y \sin{Y} \right)  \right] \right\} \,.
\label{44}
\end{equation}
Formally speaking, it contains the information about the axion field (see the last term in this formula), however, we include this term into the tensor $T^{(\rm U)}_{ik}$, since it originates from the Lagrange multiplier $\lambda$ (\ref{34}). The second contribution to the total stress-energy tensor given by
\begin{equation}
T^{(\rm A)}_{ik} = \Psi^2_0 {\dot{\phi}}^2 \left[U_iU_k {-} \frac12 g_{ik} \right] {+}
\label{45}
\end{equation}
$$
+g_{ik} m^2_A \Psi^2_0 \left(\frac{\Phi_*}{2\pi} \right)^2\left(1{-}\cos{Y} \right)  {+}
$$
$$
+ \frac{m^2_A \Psi^2_0}{2\pi^2} g_{ik} \left(\Theta {+} \frac{d}{dt} \right) \left[\Phi_* \frac{d \Phi_{*}}{d\Theta}\left(\frac12 Y \sin{Y}  {-} 1{+}\cos{Y} \right)\right]
$$
is attributed to the axion field. The last term
$$
T^{(\rm MA)}_{ik} = \cos{Y}\left[\frac14 g_{ik} F_{pq}F^{pq} - F_{ip}F_k^{\ p}\right] +
$$
$$
+ \frac{1}{16\pi} g_{ik} \left(\Theta + \frac{d}{dt} \right) \left\{\frac{d \Phi_{*}}{d\Theta} \left[ F^{*}_{pq}F^{pq} \left(Y \cos{Y} {-} \sin{Y}\right) - \right. \right.
$$
\begin{equation}
\left. \left.- \left(\frac{2\pi}{\Phi_*}\right) Y \sin{Y} F_{pq}F^{pq} \right] \right\}
\label{46}
\end{equation}
is indicated by the letters $MA$ and describes the Maxwell field coupled to the nonlinear axion field. Clearly, the trace of this tensor is nonvanishing. When $Y=0$ we obtain the standard traceless stress-energy tensor of the electromagnetic field in vacuum.

In order to facilitate the formulation of the reduced equations of the gravity field we visualize the nonvanishing components of the stress-energy tensors presented above.

Mention should be made that due to the structure of the aether vector field, $U^i=\delta^i_0$, three spatial components of the term ${T^{(\rm U)}}^i_{k}$ coincide:
\begin{equation}
 {T^{(\rm U)}}^1_{1}= {T^{(\rm U)}}^2_{2} = {T^{(\rm U)}}^3_{3} = - C_2 \left(\dot{\Theta} + \frac12 \Theta^2 \right) \,,
\label{47}
\end{equation}
thus emphasizing that the aether flow is spatially isotropic, and the aether pressure is of the Pascal form. The last nonvanishing component has more sophisticated form:
$$
{T^{(\rm U)}}^0_{0} = - \frac12 C_2  \Theta^2   +
$$
$$
+ \frac{\kappa}{8\pi}  \frac{d}{dt}\left\{\left(\frac{\Phi_*}{2\pi}\right)\frac{d\Phi_*}{d\Theta}\left\{\frac{2B_0^2}{a^2 b^2} \left[2 \tan{Y} \left(\sin{Y}-Y \cos{Y} \right) + \right. \right. \right.
$$
$$
\left. \left. \left. + Y \sin{Y} \left(\left(\frac{2\pi}{\Phi_*}\right)^2-  \tan^2{Y}\right) \right] + \right. \right.
$$
\begin{equation}
\left. \left. + 8 m^2_A \Psi^2_0  \left(1{-}\cos{Y} {-} \frac12 Y \sin{Y} \right)  \right\} \right\} \,.
\label{48}
\end{equation}
Similarly, three spatial components of the tensor ${T^{(\rm A)}}^i_{k}$ coincide
\begin{equation}
{T^{(\rm A)}}^1_{1} = {T^{(\rm A)}}^2_{2} = {T^{(\rm A)}}^3_{3}=
\label{49}
\end{equation}
$$
- \frac12 \Psi^2_0 {\dot{\phi}}^2  {+} m^2_A \Psi^2_0 \left(\frac{\Phi_*}{2\pi} \right)^2\left(1{-}\cos{Y} \right)  {+}
$$
$$
+ \frac{m^2_A \Psi^2_0}{2\pi^2} \left(\Theta {+} \frac{d}{dt} \right) \left\{\Phi_* \frac{d \Phi_{*}}{d\Theta}\left[\frac12 Y \sin{Y}  {-} 1{+}\cos{Y} \right]\right\} \,,
$$
demonstrating the fact that the pressure of the axion field also is of the Pascal type. The energy density of the axion field can be written as follows:
\begin{equation}
{T^{(\rm A)}}^0_{0} = \frac12 \Psi^2_0 {\dot{\phi}}^2    {+}  m^2_A \Psi^2_0 \left(\frac{\Phi_*}{2\pi} \right)^2\left(1{-}\cos{Y} \right)  {+}
\label{50}
\end{equation}
$$
+ \frac{m^2_A \Psi^2_0}{2\pi^2} \left(\Theta {+} \frac{d}{dt} \right) \left\{\Phi_* \frac{d \Phi_{*}}{d\Theta}\left[\frac12 Y \sin{Y}  {-} 1{+}\cos{Y} \right]\right\} \,.
$$
The last contribution, ${T^{(\rm MA)}}^i_k$, is spatially anisotropic because of the presence of the magnetic field. Since the axis $0z$ is selected, we obtain that the components of the anisotropic pressure in the plane $x0y$
coincide
$$
{T^{(\rm MA)}}^1_{1}  = {T^{(\rm MA)}}^2_{2}= {-} \frac{B^2_0}{2a^2 b^2} \cos{Y} \left[1{+} \left(\frac{\Phi_*}{2\pi} \right)^2 \tan^2{Y}\right] {+}
$$
$$
+ \frac{1}{8\pi} \left(\Theta + \frac{d}{dt} \right) \left\{\frac{B_0^2}{a^2 b^2} \left(\frac{\Phi_*}{2\pi}\right)\frac{d \Phi_{*}}{d\Theta} \sin{Y} \times \right.
$$
\begin{equation}
\left. \times \left[ 2  \left(Y  {-} \tan{Y} \right) - Y  \left[\left(\frac{2\pi}{\Phi_*}\right)^2-  \tan^2{Y}\right] \right] \right\} \,.
\label{51}
\end{equation}
As well,  the components ${T^{(\rm MA)}}^0_0$ and ${T^{(\rm MA)}}^3_3$ coincide:
$$
{T^{(\rm MA)}}^0_0 = {T^{(\rm MA)}}^3_3 =  \frac{B^2_0}{2a^2 b^2} \cos{Y} \left[1{+} \left(\frac{\Phi_*}{2\pi} \right)^2 \tan^2{Y}\right] {+}
$$
$$
+ \frac{1}{8\pi} \left(\Theta + \frac{d}{dt} \right) \left\{\frac{B_0^2}{a^2 b^2} \left(\frac{\Phi_*}{2\pi}\right)\frac{d \Phi_{*}}{d\Theta} \sin{Y} \times \right.
$$
\begin{equation}
\left. \times \left[ 2  \left(Y  {-} \tan{Y} \right) - Y  \left[\left(\frac{2\pi}{\Phi_*}\right)^2-  \tan^2{Y}\right] \right] \right\} \,.
\label{52}
\end{equation}

5\subsection{The model with axion field in the equilibrium state }

\subsubsection{First submodel with $C_2 \neq - \frac23$: The key equations and asymptotic behavior of the solutions}

We indicate the states of the axionic system as the equilibrium ones, when $\phi {=} n \Phi_*$, since this case corresponds to one of the minima of the potential of the axion field (\ref{13}). Clearly, now $Y=2\pi n$, thus $\sin{Y}=0$ and the right-hand side of the equation (\ref{43}) vanishes. The equation for the axion field converts into
\begin{equation}
\ddot{\Phi_*} + \Theta \dot{\Phi_*} = 0 \,,
\label{53}
\end{equation}
and admits the first integral
\begin{equation}
\dot{\Phi}_*(t) = \frac{\rm const}{a(t)b(t)c(t)} \,.
\label{54}
\end{equation}
For this solution $E(t)=0$, the electric field can not be generated, and we are faced with the anisotropic cosmological model with dynamic aether and pure magnetic field. It can be considered as an extension of the known model with the pure magnetic field.
We see that now the gravity field equations take the following form:
$$
\left[\frac{\dot{a}}{a} \frac{\dot{b}}{b} {+} \frac{\dot{a}}{a} \frac{\dot{c}}{c} {+} \frac{\dot{b}}{b} \frac{\dot{c}}{c} \right] - \Lambda + \frac12 C_2 \Theta^2  =
$$
\begin{equation}
= \frac{\kappa}{2} \left[\left(\frac{\Psi^2_0 K}{abc}\right)^2  {+} \left(\frac{B_0}{a b}\right)^2 \right]  \,,
\label{55}
\end{equation}
$$
\left[\frac{\ddot{b}}{b} +  \frac{\ddot{c}}{c} + \frac{\dot{b}}{b} \frac{\dot{c}}{c} \right]  - \Lambda  + C_2 \left(\dot{\Theta}  + \frac12 \Theta^2  \right) =
 $$
 \begin{equation}
 = - \frac{\kappa}{2} \left[\left(\frac{\Psi^2_0 K}{abc}\right)^2  + \left(\frac{B_0}{a b}\right)^2 \right]  \,,
\label{56}
\end{equation}
$$
\left[\frac{\ddot{a}}{a} +  \frac{\ddot{c}}{c} + \frac{\dot{a}}{a} \frac{\dot{c}}{c}\right]  - \Lambda  + C_2 \left(\dot{\Theta}  + \frac12 \Theta^2  \right) =
 $$
 \begin{equation}
 = - \frac{\kappa}{2} \left[\left(\frac{\Psi^2_0 K}{abc}\right)^2  + \left(\frac{B_0}{a b}\right)^2\right] \,,
\label{57}
\end{equation}
$$
\left[\frac{\ddot{b}}{b} +  \frac{\ddot{a}}{a} + \frac{\dot{b}}{b} \frac{\dot{a}}{a}\right]  - \Lambda  + C_2 \left(\dot{\Theta}  + \frac12 \Theta^2  \right) =
$$
\begin{equation}
  =-\frac{\kappa}{2} \left[\left(\frac{\Psi^2_0 K}{abc}\right)^2  - \left(\frac{B_0}{a b}\right)^2 \right] \,.
\label{58}
\end{equation}
The term $\left(\dot{\Theta}  + \frac12 \Theta^2  \right)$, which enters the gravity field equations can be decoded as
$$
\dot{\Theta}  + \frac12 \Theta^2  = \left(\frac{\ddot{a}}{a} +\frac{\ddot{b}}{b} +\frac{\ddot{c}}{c}\right) +
$$
\begin{equation}
 + \left(\frac{\dot{a}}{a}\frac{\dot{b}}{b} + \frac{\dot{a}}{a}\frac{\dot{c}}{c} + \frac{\dot{b}}{b}\frac{\dot{c}}{c} \right) - \frac12 \left(\frac{{\dot{a}}^2}{a^2} + \frac{{\dot{b}}^2}{b^2} +\frac{{\dot{c}}^2}{c^2} \right) \,.
\label{59}
\end{equation}
We see that four equations (\ref{55})-(\ref{58}) contain three metric coefficients $a(t)$, $b(t)$, $c(t)$ and their derivatives only. Since the equations for the electromagnetic and aether vector fields are already solved, and the consequence of the first integral of the axion field equation is inserted into the gravity field equations, then the Bianchi identity is the trivial consequence of (\ref{55})-(\ref{58}). In other words, one of the four equations can be omitted. Also, one can see that the magnetic field $B_0$ is the only spatially anisotropic source of the gravity field, and we can consider the metric with $a(t)=b(t)$ thus assuming that there exists local rotation invariance with respect to the axis $0z$. Finally, we assume that the basic state of the equilibrium axion field is associated with the constant value of the function $\Phi_*$, which yields $K=0$. Then we obtain the following two equations for two unknown functions $H(t)$ and $\Theta(t)$:
\begin{equation}
2H \Theta -3 H^2   + \frac12 C_2 \Theta^2  = \Lambda +
\frac{\kappa B^2_0}{2 a^4}  \,,
\label{60}
\end{equation}
\begin{equation}
2 \dot{H} + 3H^2    + C_2 \left(\dot{\Theta}  + \frac12 \Theta^2  \right) = \Lambda + \frac{\kappa B^2_0}{2 a^4} \,.
\label{61}
\end{equation}
The unknown metric functions $a(t)$ and $c(t)$ enter the functions $H(t)$ and $\Theta(t)$ as follows:
\begin{equation}
H(t) \equiv \frac{\dot{a}}{a} \,, \quad \Theta(t) = 2 \frac{\dot{a}}{a} + \frac{\dot{c}}{c} \,.
\label{62}
\end{equation}
As a consequence of (\ref{60}), (\ref{61}), the functions $H(t)$ and $\Theta(t)$ are linked by the relationship
\begin{equation}
\frac{d}{dt}\left(H+ \frac12 C_2 \Theta\right) = H (\Theta - 3H) \,.
\label{63}
\end{equation}
In the asymptotic regime, when $t \to \infty $, $a \to \infty$, $\dot{H} \to 0$, $\dot{\Theta} \to 0$, we obtain from (\ref{63}) and (\ref{60} )that
\begin{equation}
\Theta(t \to \infty) \to 3H_{\infty} \,,
\label{64}
\end{equation}
\begin{equation}
 H(t \to \infty) \to H_{\infty} \equiv \sqrt{\frac{\Lambda}{3 \Gamma}} \,, \quad \Gamma = 1 + \frac32 C_2 \neq 0 \,.
\label{65}
\end{equation}
These asymptotic formulas  are valid,  first, when, $\Lambda>0$ and $\Gamma>0$ (de Sitter type asymptote); second, when $\Lambda<0$ and $\Gamma<0$ (anti-de Sitter type asymptote).
This means that asymptotically
\begin{equation}
\frac{a(t)}{a_*} \Rightarrow \ e^{H_{\infty}t} \ \Leftarrow \ \frac{c(t)}{c_*} \,,
\label{66}
\end{equation}
and we deal with the isotropization of the Universe. Clearly, the asymptotic de Sitter(anti-de Sitter) regime is provided both: by the cosmological constant $\Lambda$ and by the dynamic aether via the coupling constant $C_2$. When $C_2=-\frac23$, we deal with the special case.

\subsubsection{On the solutions to the key equations for the case $C_2 \neq - \frac23$}

One can reduce the system of equations (\ref{60}), (\ref{61}) to one key equation for $\Theta$ using the following scheme. First, we introduce the variable $x$ as follows:
\begin{equation}
x=\frac{a(t)}{a(t_0)} \,, \quad \frac{d}{dt} = xH(x)\frac{d}{dx} \,.
\label{67}
\end{equation}
Second, we extract the function $H(x)$ from the quadratic equation (\ref{60}):
\begin{equation}
H(x) {=}  \frac13 \Theta \pm \sqrt{\frac19 \Theta^2 \left(1{+} \frac32C_2\right) {-} \left(\frac{\Lambda}{3} {+} \frac{\kappa B_0^2}{6a^4(t_0) x^4} \right)} \,.
\label{68}
\end{equation}
Third, we put this $H(x)$ into (\ref{61}) transformed as follows:
\begin{equation}
x\frac{d}{dx}\left[H(x)+ \frac12 C_2 \Theta(x)\right] =  \Theta(x) - 3H(x) \,.
\label{69}
\end{equation}
We obtain the nonlinear equation for the function $\Theta(x)$, which can be written in the form
\begin{equation}
x\frac{d}{dx}\left[\Theta(x) \pm {\cal F}\right] =  \mp 3 {\cal F} \,,
\label{70}
\end{equation}
where
\begin{equation}
{\cal F} \equiv \frac{3}{\left(1{+}\frac32 C_2 \right)} \sqrt{\frac19 \Theta^2 \left(1{+} \frac32C_2\right) {-} \left(\frac{\Lambda}{3} {+} \frac{\kappa B_0^2}{6a^4(t_0) x^4} \right)} \,.
\label{71}
\end{equation}
In fact, this equation is the key equation of the model, since its solution $\Theta(x)$ opens the process of subsequent finding of the functions $H(x)$, $a(t)$, $c(t)$.
This equation is nonlinear and requires the numerical analysis, but here we attract the attention to one exact solution to this key equation, which appears if $\Lambda=0$.
Clearly, there exists the solution of the form
\begin{equation}
\Theta(x) = \frac{\nu}{x^2}\,, \quad H(x) = \frac{\nu (1+C_2)}{x^2}\,,
\label{72}
\end{equation}
where the parameter $\nu$ depends on $B_0$ and $C_2$ as follows:
\begin{equation}
\nu = \sqrt{\frac{\kappa  B_0^2}{2a^4(t_0) \left(1+\frac32 C_2 \right)(1+2C_2)}} \,.
\label{73}
\end{equation}
The solution is real, if $-\frac23 <C_2 <-\frac12$.
The first metric coefficient $a(t)$ can be found from the relation
\begin{equation}
t{-}t_0 = \int_1^{\frac{a(t)}{a(t_0)}} \frac{dx}{xH(x)} = \frac{1}{2\nu(1{+}C_2)} \left[\left(\frac{a(t)}{a(t_0)} \right)^2 {-}1 \right] \,,
\label{74}
\end{equation}
which gives
\begin{equation}
a(t) = a(t_0) \sqrt{1+ 2\nu(1+C_2)( t-t_0)} \,.
\label{75}
\end{equation}
The corresponding asymptote is characterized by the law
\begin{equation}
a(t \to \infty) \propto t^{\frac12}  \,.
\label{76}
\end{equation}
The second metric coefficient $c(t)$ can be found from the relationships
\begin{equation}
\frac{\dot{c}}{c} = \Theta {-}2H = {-} \frac{\nu (1{+}2C_2)}{x^2} = {-} \frac{\nu (1{+}2C_2)}{1{+} 2\nu(1{+}C_2)( t{-}t_0)} \,.
\label{77}
\end{equation}
We obtain now the law
\begin{equation}
c(t) = \frac{c(t_0)}{1 + 2\nu (1+C_2)(t-t_0)} \,,
\label{78}
\end{equation}
with the asymptote
\begin{equation}
c(t) \propto \frac{1}{t} \,.
\label{79}
\end{equation}
In other words, the Universe described by this model expands in the plane $x0y$ and compresses in the direction $0z$ so that asymptotically $abc \to const$.

\subsubsection{The special case $C_2 = - \frac23$}

For this special aether configuration the equation (\ref{63}) can be rewritten as
\begin{equation}
\frac{d}{dt} \left(\Theta-3H \right) = -3H \left(\Theta-3H \right) \,.
\label{80}
\end{equation}
The solution to this equation is
\begin{equation}
\Theta-3H = \frac{\rm const}{a^3} \,,
\label{81}
\end{equation}
and the equation (\ref{60}) yields
\begin{equation}
\Lambda + \frac{\kappa B^2_0}{2 a^4} + \frac{{(\rm const)}^2}{a^6} =0 \,.
\label{82}
\end{equation}
This equation has no real solutions, when $\Lambda \geq 0$. In the anti - de Sitter model, when $\Lambda <0$, the real solution for $a(t)$ exists, and it is the constant solution $a(t)= a_*$.
Then $H(t)=0$, and
\begin{equation}
\Theta = \Theta_* \equiv \frac{\rm const}{a^3_*} \,.
\label{83}
\end{equation}
This means that in the plane orthogonal to the magnetic field the Universe does not expand, and the expansion/compression in the $0z$ direction takes place with the exponential law
\begin{equation}
c(t)= c_* e^{\Theta_*  t} \,.
\label{84}
\end{equation}
When the constant of integration vanishes, ${\rm const}=0$, we deal with an anisotropic stationary Universe, which is described by the parameters
\begin{equation}
H=0 \,, \quad \Theta =0 \,, \quad a_* =  \left(\frac{\kappa B^2_0}{2|\Lambda|} \right)^{\frac14}, \quad  c_* = const \,.
\label{85}
\end{equation}
In other words, the size of the stationary Universe in the plane $x0y$ is predetermined by the ratio between the amplitude of the magnetic field and the modulus of the negative cosmological constant, while the size in the direction $0z$ is given by an arbitrary constant.

\subsection{The second submodel: The axion system is in the state of the dynamic equilibrium}

The right-hand side of the equation (\ref{43}) vanishes, when the function $Y$ satisfies the equation
\begin{equation}
\tan^2{Y} = \left(\frac{2\pi}{\Phi_*}\right)^2 -\frac{2}{B_0^2} \left(B^2_0+ \Psi^2_0 m^2_A a^2 b^2 \right) \,.
\label{86}
\end{equation}
We indicate this situation as the dynamic equilibrium in contrast to the standard equilibrium state of the axionic dark matter. The solution to this equation
\begin{equation}
\frac{2\pi \phi(t)}{\Phi_*(t)}= \pi k \pm \arctan{ \sqrt{\left(\frac{2\pi}{\Phi_*}\right)^2 {-} 2 \left(1 {+} \frac{\Psi^2_0 m^2_A a^2 b^2}{B_0^2} \right)}}
\label{87}
\end{equation}
has to be supplemented by the requirement
\begin{equation}
\dot{\phi}=\frac{K}{abc} \,,
\label{88}
\end{equation}
where $K$ is an integration constant. If, as in the previous case, we put $a(t)=b(t)$, and $K=0$, obtaining $\phi(t) = \phi(t_0)$, the guiding  function $\Phi_*(t)$ will be found as the solution to the transcendent equation, which can be represented in the following convenient form:
\begin{equation}
a^4(Y)  = {\tilde{a}}^4  \left[\left(\frac{Y}{\phi(t_0)}\right)^2 - 2 - \tan^2{Y} \right]  \,,
\label{898}
\end{equation}
\begin{equation}
\Phi_*(t) = \frac{2\pi \phi(t_0)}{Y(t)} \,, \quad  {\tilde{a}}^4  = \frac{B_0^2}{2 \Psi^2_0 m^2_A} \,.
\label{897}
\end{equation}
The real solution exists, when $\left(\frac{Y}{\phi(t_0)}\right)^2 > 2 + \tan^2{Y}$; the corresponding time interval is non-empty, when this inequality holds for $t=t_0$, as well as, the derivative of the right-hand side of (\ref{898}) at the starting point is positive, i.e., when
$$
|\phi(t_0)| < \frac{|Y(t_0) \cos{Y(t_0)}|}{\sqrt{1+ \cos^2{Y(t_0)}}} \,,
$$
\begin{equation}
|\phi(t_0)| < \sqrt{\left| \frac{Y(t_0) \cos^3{Y(t_0)}|}{\sin{Y(t_0)}}\right|} \,.
\label{811}
\end{equation}
The dynamic equilibrium can not exist during the whole interval of the cosmological expansion $t_0<t<\infty$; it can take place up to the critical moment $t=t_{C}$, which is defined via the critical value  $Y(t_C)$, indicated as $Y_{C}\neq 0$. This critical parameter can be found from the condition that the scale factor reaches the maximum, and it satisfies the transcendent equation
\begin{equation}
Y_{C} \cos^3{Y_C}= \phi^2(t_0) \sin{Y_{C}} \,.
\label{90}
\end{equation}
The corresponding maximum value of the Universe radius is given by
$$
a_{C} \equiv a(t_C) =
$$
\begin{equation}
= \sqrt{\frac{|B_0|}{\sqrt2 \Psi_0 m_{A}}} \left\{\frac{Y^2_C}{\phi^4(t_0)}\left[\phi^2(t_0){-}\cos^4{Y_C} \right]{-}2 \right\}^{\frac14} \,,
\label{91}
\end{equation}
At the critical moment of the cosmological time $t_C$ the dynamic equilibrium is assumed to be destroyed due to the cosmological expansion, and the model under discussion becomes invalid. The detailed analysis of the Universe evolution in the time interval $t_0<t<t_{C}$ can be fulfilled only numerically; we hope to do this work in the nearest future.

We would like to emphasize the following detail: the "catastrophic" moment of the cosmological time, $t_{\times}$, for which $\tan(Y(t_{\times}))=\infty$ and the model becomes singular, satisfies the condition $t_{\times} > t_C$, thus, we do not take into account such singularity.

\subsection{On the stability of the axionic equilibrium state}

When we speak about the problem of the axionic state stability, we keep in mind two questions. The first question is connected with vanishing of the $\cos{Y}$: whether the axion field can reach the value $\phi = \pm \frac14 \Phi_*$? If yes, we obtain the catastrophic situation with $E=\infty$, $F_{mn}F^{mn}=\infty$, etc. If these quantities take finite values at the finite moments of the cosmological time, the second question arises connected with the asymptotic stability of the solutions to the equation of the axion field.

\subsubsection{When can the catastrophic regime be realized?}

The catastrophic regime takes place, when during the finite time interval the function $\cos{Y}$ reaches zero value. For this special behavior of the axion field we can suppose that the scale factor changes insignificantly, and we can put $a(t) \to a_{(\rm m)}$, and $\Theta \to 0$. Also, we assume that during this short interval of time the guiding function remains constant, and we choose $\left(\frac{2\pi}{\Phi_*}\right)^2 = 1 + \frac{2m^2_A \Psi^2_0 a^4_{(\rm m})}{B^2_0}$. Then after the time rescaling $\tau = t \left(\frac{B_0}{\Psi_0 a^2_{(\rm m)}}\right)$ we obtain the following dimensionless  equation:
\begin{equation}
\frac{d^2 Y}{d \tau^2} = - \frac{\sin{Y}}{2\cos^2{Y}} \,.
\label{catas1}
\end{equation}
Using the multiplier $\frac{d Y}{d \tau}$ we obtain after the standard procedure the first integral
\begin{equation}
\left(\frac{d Y}{d \tau}\right)^2(\tau) = \left[\left(\frac{d Y}{d \tau}\right)^2(\tau_0) + \frac{1}{\cos{Y(\tau_0)}}\right] - \frac{1}{\cos{Y}} \,,
\label{catas2}
\end{equation}
where the moment $\tau_0$ is chosen so that the value of the derivative $\left(\frac{d Y}{d \tau}\right)(\tau_0)$ is finite.

\noindent
1) If $Y(\tau)$, and $Y(\tau_0)$ also, belong to the interval $(-\frac{\pi}{2},\frac{\pi}{2})$, so that $\cos{Y}>0$, there exists some moment $\tau = \tau_{(\rm m)}$, for which the right-hand  side of the equation (\ref{catas2}) vanishes. If $\frac{1}{\cos{Y}}$ continues to grow, the derivative becomes imaginary.  Thus, at this moment $\left(\frac{d Y}{d \tau}\right)(\tau_{(\rm m )})=0$, and we deal with the maximum of the function $Y(\tau)$, and see that $Y$ is bounded and $Y_{(\rm max)}<\frac{\pi}{2}$. The catastrophic regime can not be realized in this self-closed domain of the integral curves.

\noindent
2) If $Y(\tau)$ belongs to the interval $(\frac{\pi}{2}, \frac{3\pi}{2})$, so that $\cos{Y}<0$, the right-hand  side of the equation (\ref{catas2}) becomes unbounded and can reach arbitrary value. In this domain the catastrophic regime is possible.

The same results can be obtained based on the theory of autonomous dynamic systems; indeed, the equation (\ref{catas1}) can be rewritten as
\begin{equation}
\frac{dY}{d \tau} = X \,, \quad \frac{dX}{d \tau} = - \frac{\sin{Y}}{2\cos^2{Y}} \,.
\label{catas10}
\end{equation}
In fact, this set of equations represents the modified physical pendulum; the singular points remain the same (the knots at $Y=2\pi m$ and saddle points at $Y= \pi (2m+1)$); the term $\frac{1}{2\cos^2{Y}}$ adds the vertical separatrices $Y=\frac{\pi}{2}+\pi k$. In the domains containing the knots, all the integral curves are closed and finite; in the domains containing the saddle points the integral curves are disclosed, unbounded, so the catastrophic regime is possible.

Clearly, the  model, which we used for interpretation, is truncated; we assume that the complete model will be analyzed in the next work.

\subsubsection{Asymptotic stability of the solutions near the equilibrium state}

Let us consider homogeneous variations of the axion field $\phi(t) \to  n \Phi_*(t) + \psi(t)$, where the quantity $|\psi|$ is small.
In the first order approximation with respect to $\psi$ the equation (\ref{43}) can be written as follows:
\begin{equation}
\ddot{\psi} + \Theta \dot{\psi} =
\psi \left\{\frac{B^2_0}{2\Psi^2_0 a^2 b^2}\left[\left(\frac{2\pi}{\Phi_*}\right)^2 {-} 2 \right] {-} m^2_A \right\} \,.
\label{92}
\end{equation}
We assume that the functions $a(t)$, $b(t)$, $\Phi_*(t)$ are found from the equations of the leading order.
With the replacement
\begin{equation}
\psi(t) = \frac{Z(t)}{\sqrt{abc}} \,,
\label{93}
\end{equation}
we reduce the equation (\ref{92}) to the form
\begin{equation}
\ddot{Z} + {\cal J}(t) Z = 0 \,,
\label{94}
\end{equation}
where  the quantity
\begin{equation}
{\cal J} {=}  m^2_A {-} \frac12 \left(\dot{\Theta} {+} \frac12 \Theta^2 \right) {+} \frac{B^2_0}{2\Psi^2_0 a^2 b^2}\left[2{-}\left(\frac{2\pi}{\Phi_*}\right)^2 \right]
\label{95}
\end{equation}
plays the role of the square of the effective frequency. In the asymptotic regime, when $\Theta \to \sqrt{\frac{3\Lambda}{\Gamma}}$, the solution for $\psi$ depends on the sign of the parameter
\begin{equation}
{\cal J}_{\infty} {=}  m^2_A {-}  \frac{3\Lambda}{4\Gamma} \,.
\label{96}
\end{equation}
When $m_A > \sqrt{\frac{3\Lambda}{4\Gamma}}$, the solution for $\psi$ is of the form
\begin{equation}
\psi {=} \frac{1}{\sqrt{abc}} \left[\alpha_1 \cos{\mu t} {+} \alpha_2 \sin{\mu t} \right], \quad \mu {=} \sqrt{ m^2_A {-}  \frac{3\Lambda}{4\Gamma}} .
\label{97}
\end{equation}
Fluctuations of this type vanish with time.

\noindent
When $m_A < \sqrt{\frac{3\Lambda}{4\Gamma}}$, the corresponding asymptotic solutions have the form
\begin{equation}
\psi = \frac{1}{\sqrt{abc}} \left[\beta_1 e^{\xi t} + \beta_2 e^{-\xi t}\right] \,, \quad \xi = \sqrt{- m^2_A {+}  \frac{3\Lambda}{4\Gamma}} \,.
\label{98}
\end{equation}
Keeping in mind that in the asymptotic regime $abc \propto e^{\sqrt{\frac{3\Lambda}{\Gamma}}t}$, we see that the first mode with $\beta_1$ in front behaves as
\begin{equation}
\psi \propto \exp{\left\{\left[\sqrt{\frac{3\Lambda}{4\Gamma}- m^2_A} - \sqrt{\frac{3\Lambda}{4\Gamma}}\right]t \right\}} \to 0 \,.
\label{99}
\end{equation}
Finally, when $m_A = \sqrt{\frac{3\Lambda}{4\Gamma}}$ and the leading order term is
\begin{equation}
\psi \propto  \frac{t}{\sqrt{abc}} \to t e^{- \sqrt{\frac{3\Lambda}{4\Gamma}}t} \,,
\label{100}
\end{equation}
this mode also vanishes asymptotically. In other words, the equilibrium state of the axionic dark matter is asymptotically stable with respect to homogeneous fluctuations.
Complete stability analysis is under preparation, but is beyond the frame of this work.

\section{Discussion and conclusions}

1. In the framework of the Einstein-Maxwell-axion-aether theory we have formulated the model, into which the pseudoscalar (axion) field enters nonlinearly, and the electromagnetic and the unit vector fields appear in the linear form. This approach is based on the idea that the equations of electromagnetism are invariant with respect to the rotations in the plane coordinated by the electric and magnetic fields \cite{Jackson}. This symmetry hinted us to modify non-linearly (sin-type non-linearity) the action functional of the axion electrodynamics, and we obtained that in addition to the sin-type modification of the axion-photon term, we have found out the cosine-type modification of the dilaton-photon term. In other words, working with only one (axion) field, one can effectively introduce the dilaton terms. This means that the nonlinear axion-photon interactions  can mimic the dilaton-photon coupling, or equivalently, the dilaton field can be considered as the field induced by the axion field.

2. The behavior of the axionic dark matter is regulated by the modified periodic potential (\ref{13}), in which the guiding function $\Phi_*$ plays the role of a vacuum average value of the pseudoscalar field. The same guiding function is used in (\ref{11}), though one could introduce another guiding function, say $\Phi_{**}$. We keep in mind the arguments of the Occam's razor  concept, and put $\Phi_{**}=\Phi_*$. The bonus of this idea is the following. In addition to the standard equilibrium states, for which $\phi=n \Phi_*$ and thus the axion potential and its derivative vanish ($n$ is an integer), we got the opportunity to introduce the state of a new type, namely, the dynamic equilibrium state. Mathematically, this state is defined by (\ref{86}), (\ref{87}); from the physical point of view it relates to the case, when the axion potential and its derivative are not equal to zero, but the axionic contribution to the equation (\ref{43}) is compensated by the electromagnetic one.
In principle, there exists a mixed equilibrium state, when both requirements are valid simultaneously (see, e.g., (\ref{143})), however, such an equilibrium can take place  only in stationary models with constant scale factors.

3. In the model under discussion the sin-type axion multiplier, appearing in front of the term $\frac14 F_{ik}F^{*ik}$, generates the cosine-type multiplier in front of the term $\frac14 F_{ik}F^{ik}$. This dilatonic type term in the Lagrangian, which has the form $\frac14 \cos{\left(\frac{2\pi \phi}{\Phi_*}\right)}F_{ik}F^{*ik}$, revives the idea of an anomaly, which could appear, when the argument of the cosine tends to $\frac{\pi}{2}+ \pi m$. In particular, in the Bianchi-I anisotropic homogeneous cosmological model with initial magnetic field we see that the axionically induced electric field (\ref{40}) and the corresponding sources of the gravity field (see, e.g., (\ref{51})) contain the function $\cos^{-1}{\left(\frac{2\pi \phi}{\Phi_*}\right)}$, which can tend to infinity when $\phi \to \frac14 \Phi_*$.
According to the illustration discussed in Section IIID1, if the starting state of the axionic dark matter is near the equilibrium point, related to the minimum of the axion potential, the catastrophic regime can not be realized, though the axionically induced electric field can reach rather big values. If the system starts to evolve from the state related to the maximum of the axion field potential, the anomalous grow of the state functions is inevitable. Such a model might be useful in the theory of anisotropic cosmological inflation.

4. What is the role of the dynamic aether in the cosmological evolution of the axion-photon system? The obtained results contain only one Jacobson's parameter, $C_2$. The parameter $C_3$ is assumed to be equal to $-C_1$, providing the speed of gravity waves in the aether coincides with the speed of light \cite{GRB17}. The  parameters $C_1$, $C_4$ remain hidden in the model with the chosen spacetime symmetry. The parameter $C_2$ enters the asymptotic value of the Hubble function $H_{\infty}= \sqrt{\frac{\lambda}{3 \Gamma}}$, $\Gamma=1{+}\frac32 C_2$, thus predetermining the rate of the Universe expansion on the final stage of evolution. The critical value of this parameter $C_2=-\frac23$ relates to the  model, for which the size of the Universe in the plane orthogonal to the magnetic field is fixed. The main channel of influence of the dynamic aether on the axionic dark matter is connected with the covariant divergence of the velocity four-vector of the aether, $\Theta=\nabla_k U^k$. We introduced this scalar into the guiding function $\Phi_*(\Theta)$, which enters the potential of the axion field (\ref{13}), describes the vacuum average value of the pseudoscalar field and predetermines the scale of the equilibrium levels for the axion field, $\phi = n \Phi_*$.

5. The presented model opens up a wide range of opportunities for modeling of the physical phenomena in the expanding Universe. In this paper we formulated the key equations of the Bianchi-I model with local rotation symmetry, and discussed examples of exact solutions; the detailed numerical analysis is in our nearest plans. For instance, generally, the obtained key equation (\ref{70}) with (\ref{71}) requires numerical modeling with wide choice of the guiding parameters $\Lambda$, $C_2$, $B_0$; however, for the special submodel with $\Lambda=0$ and $\Phi_*=const$, we have found the exact solution in the analytic form (\ref{72}) with (\ref{73}), which allows us to reconstruct immediately the scale factors $a(t)$ (\ref{75}), and $c(t)$ (\ref{78}).

\acknowledgments{The work was supported by the Russian Foundation for Basic Research (Grant N 20-52-05009)}


\section*{References}

\begin{thebibliography}{99}

\bibitem{1} J. Plebanski, {\it Lectures on Nonlinear Electrodynamics} (Nordita, Copenhagen, 1970).

\bibitem{2} G. Boillat, Nonlinear electrodynamics - Lagrangians and equations of motion, J. Math. Phys. {\bf 11}, 3, 941-951 (1970).

\bibitem{3} H.K. Avetissian, {\it Relativistic Nonlinear Electrodynamics: The QED Vacuum and Matter in super-strong Radiation Fields} (Springer Series on Atomic, Optical, and Plasma Physics,
Springer, New York, 2016).

\bibitem{5} K.A. Bronnikov, Regular magnetic black holes and monopoles from nonlinear electrodynamics, Phys. Rev. D {\bf 63}, 044005 (2001).

\bibitem{6} S.I. Kruglov, Dyonic and magnetized black holes based on nonlinear electrodynamics, Eur. Phys. J. C  {\bf 80}, 250 (2020).

\bibitem{7} A.A. Chernitskii, Dyons and interactions in nonlinear (Born-Infeld) electrodynamics, JHEP {\bf 9912}, 010 (1999).

\bibitem{8} L. Hollenstein and F.S.N. Lobo, Exact solutions of f(R) gravity coupled to nonlinear electrodynamics, Phys. Rev. D {\bf 78}, 124007 (2008).

\bibitem{9} A. Sheykhi and S. Hajkhalili, Dilaton black holes coupled to nonlinear electrodynamic field, Phys. Rev. D {\bf 89},  104019 (2014).

\bibitem{10}  K.A. Bronnikov, Nonlinear electrodynamics, regular black holes and wormholes, Int. J. Mod. Phys. D {\bf 27}, 1841005 (2018).

\bibitem{11} K.E. Kunze, Primordial magnetic fields and nonlinear electrodynamics, Phys. Rev. D {\bf 77}, 023530 (2008).

\bibitem{12}  R. Garcia-Salcedo and N. Breton,  Nonlinear electrodynamics in Bianchi spacetimes,  Class. Quant. Grav. {\bf 20}, 5425-5437 (2003).

\bibitem{121} H.J. Mosquera Cuesta and G. Lambiase,  Nonlinear electrodynamics and CMB polarization, JCAP {\bf 1103},  033 (2011).

\bibitem{P1}  Y.N. Obukhov and G.F. Rubilar, Fresnel analysis of the wave propagation in nonlinear electrodynamics, Phys. Rev. D {\bf 66},   024042 (2002).

\bibitem{P2}  V.A. De Lorenci, R. Klippert, Shi-Yuan Li and J.P. Pereira, Multirefringence phenomena in nonlinear electrodynamics, Phys. Rev. D {\bf 88}, 065015 (2013).

\bibitem{13} J.R. Morris and A. Schulze-Halberg, Light wave propagation through a dilaton-Maxwell domain wall, Phys. Rev. D {\bf 92}, 8, 085026  (2015).

\bibitem{14} S.L. Liebling, Maxwell-dilaton dynamics, Phys. Rev. D {\bf 100}, 104040 (2019).

\bibitem{15} H.R. Christiansen, M.S. Cunha and M.O. Tahim, Exact solutions for a Maxwell-Kalb-Ramond action with dilaton: Localization of massless and massive modes in a sine-Gordon brane-world,
Phys. Rev. D {\bf 82}, 085023 (2010).

\bibitem{16} Ch. Charmousis, B. Gouteraux and J. Soda, Einstein-Maxwell-Dilaton theories with a Liouville potential, Phys. Rev. D {\bf 80}, 024028 (2009).

\bibitem{17} J. Morris, Generalized dilaton-Maxwell cosmic string and wall solutions, Phys. Lett. B {\bf 641},  1-5 (2006).

\bibitem{18} A. Herrera-Aguilar and O. Kechkin, Charging symmetries and linearizing potentials for Einstein-Maxwell dilaton-axion theory, Mod. Phys. Lett. A {\bf 13}, 1907-1914 (1998).

\bibitem{19} N. Breton, T. Matos and A. Garcia, Colliding plane waves in Einstein-Maxwell-dilaton fields, Phys. Rev. D {\bf 53}, 1868-1873 (1996).

\bibitem{20} S. Nojiri and S.D. Odintsov, Conformal anomaly for dilaton coupled electromagnetic field, Phys. Lett. B {\bf 426}, 29-35 (1998).

\bibitem{WTNi77} Wei-Tou Ni,  Equivalence principles and electromagnetism, Phys. Rev. Lett. {\bf 38},  301-304 (1977).

\bibitem{PQ} R.D. Peccei and H.R. Quinn,  CP conservation in the presence of instantons, Phys. Rev. Lett. {\bf 38}, 1440-1443 (1977).

\bibitem{Weinberg} S. Weinberg, A new light boson? Phys. Rev. Lett. {\bf 40}, 223-226 (1978).

\bibitem{Wilczek} F. Wilczek, Problem of strong P and T invariance in the presence of instantons, Phys. Rev. Lett. {\bf 40}, 279-282 (1978).

\bibitem{a1}  P. Sikivie, Experimental tests of the "invisible" axion, Phys. Rev. Lett. {\bf 51}, 1415-1417 (1983).

\bibitem{a2} F. Wilczek, Two applications of axion electrodynamics, Phys. Rev. Lett. {\bf 58}, 1799-1802 (1987).

\bibitem{a3} F.W. Hehl and Yu.N. Obukhov,  {\it Foundations of Classical Electrodynamics: Charge, Flux, and
Metric} (Birkh\"auser: Boston, USA, 2003).

\bibitem{a4}  A.B. Balakin and L.V. Grunskaya,  Axion electrodynamics and dark matter fingerprints in the terrestrial magnetic and electric fields,  Rept. Math. Phys. {\bf 71},
45-67 (2013).

\bibitem{a5} A.B. Balakin and Wei-Tou Ni, Anomalous character of the axion-photon coupling in a magnetic field distorted by a pp-wave gravitational background, Class. Quantum Grav. {\bf 31}, 105002 (2014).

\bibitem{a6} T.Yu. Alpin and A.B. Balakin,  The Einstein-Maxwell-aether-axion theory: Dynamo-optical anomaly in the electromagnetic response, Int. J. Mod. Phys. D  {\bf 25},
1650048 (2016).

\bibitem{a7} A.B. Balakin, Electrodynamics of a cosmic dark fluid, Symmetry {\bf 8} 56  (2016).

\bibitem{a8} E.M. Murchikova, On nonlinear classical electrodynamics with an axionic term, J. Phys. A {\bf 44}, 045401 (2011).

\bibitem{U1} T. Matos, G. Miranda, R. Sanchez-Sanchez and P. Wiederhold,
Class of Einstein - Maxwell - dilaton - axion space-times, Phys. Rev. D {\bf 79}, 124016 (2009).

\bibitem{U2} M. Azreg-Ainou, G. Clement and D.V. Gal'tsov, All extremal instantons in Einstein-Maxwell-dilaton-axion theory, Phys. Rev. D {\bf 84},  104042  (2011).

\bibitem{U3} L.A. Lopez and N. Breton,  Asymptotic singular behaviour of inhomogeneous cosmologies in Einstein-Maxwell-dilaton-axion theory, Gen. Rel. Grav. {\bf 39}, 153-166 (2007).

\bibitem{NU1} T. Dereli and R.W. Tucker, Born-Infeld axion-dilaton electrodynamics and electromagnetic confinement, Phys. Lett. B {\bf 703}, 530-535 (2011).

\bibitem{NU2} G.W Gibbons and D.A Rasheed,  SL(2,R) invariance of non-linear llectrodynamics coupled to an axion and a dilaton,
Phys. Lett. B {\bf 365}, 46-50 (1996).

\bibitem{ADM1}  L.D. Duffy, K.  van Bibber, Axions as dark matter particles, New J. Phys. {\bf 11}, 105008 (2009).

\bibitem{ADM2} M. Khlopov, {\it Fundamentals of Cosmic Particle Physics} (CISP-Springer: Cambridge, UK, 2012).

\bibitem{ADM3} A. Del Popolo, Nonbaryonic dark matter in cosmology, Int. J. Mod. Phys. D {\bf 23}, 1430005 (2014).

\bibitem{Jackson} J.D. Jackson, {\it Classical Electrodynamics}, (John Wiley and Sons, USA, 1999).

\bibitem{AE1} T. Jacobson and D. Mattingly, Gravity with a dynamical preferred frame, Phys. Rev. D {\bf 64} 024028 (2001).

\bibitem{AE2}  T. Jacobson, Einstein-aether gravity: a status report, PoSQG-Ph {\bf 020},  020 (2007).

\bibitem{AE3} T. Jacobson and D. Mattingly, Einstein-aether waves, Phys. Rev. D {\bf 70},  024003 (2004).

\bibitem{AE4} C. Heinicke, P. Baekler and F.W. Hehl,  Einstein-aether theory, violation of Lorentz invariance, and metric-affine gravity, Phys. Rev. D {\bf 72} 025012 (2005).

\bibitem{AE5} A.B. Balakin and J.P.S. Lemos,  Einstein-aether theory with a Maxwell field: General formalism, Ann. Phys. {\bf 350},  454-484 (2014).

\bibitem{AE6} C. Eling, T. Jacobson and M.C. Miller, Neutron stars in Einstein-aether theory, Phys. Rev. D {\bf 76},  042003 (2007).

\bibitem{Pref1} C.M. Will and K. Nordtvedt, Conservation laws and preferred frames in relativistic gravity. I. Preferred-frame theories and an extended PPN formalism,
Astrophys. J., {\bf 177},  757  (1972).

\bibitem{Pref2} K. Nordtvedt and C.M. Will, Conservation laws and preferred frames in relativistic gravity. II. Experimental evidence to rule out preferred-frame theories of gravity,
Astrophys. J. {\bf 177},  775-792   (1972).

\bibitem{LV1} S. Liberati, Lorentz breaking effective field theory and observational tests,
Lect. Notes Phys. {\bf 870},  297-342 (2013).

\bibitem{LV2} A. Kostelecky and M. Mewes, Electrodynamics with Lorentz-violating operators of arbitrary dimension, Phys. Rev. D {\bf 80},  015020 (2009).

\bibitem{LV3} C. L\"ammerzahl, A. Macias and H. M\"uller,  Lorentz invariance violation and charge (non-)conservation: A general theoretical
frame for extensions of the Maxwell equations,  Phys. Rev. D {\bf 71},  025007 (2005).

\bibitem{KT} S.R. de Groot, W.A. van Leeuven and Ch.G. van Weert, {\it Relativistic Kinetic Theory}, North-Holland, Amsterdam (1980).


\bibitem{MAG} D.Grasso and H.R. Rubinstein, Magnetic fields in the early Universe, Phys. Rept., {\bf 348}, pp. 163-266 (2001).

\bibitem{Vlasov} A.B. Balakin, R.K. Muharlyamov and A.E. Zayats, Axion-induced oscillations of cooperative electric field in a cosmic magneto-active plasma,
Eur. Phys. J. D {\bf 68}, 159 (2014).


\bibitem{Equilib1} A.B. Balakin and A.F. Shakirzyanov, Axionic extension of the Einstein-aether theory: How does dynamic aether regulate the state of axionic dark matter? Physics of the Dark Universe, {\bf 24}, 100283 ( 2019).

\bibitem{Equilib2} A.B. Balakin and D.E. Groshev, New application of the Killing vector field formalism: modified periodic potential and two-level profiles of the axionic dark matter distribution,  Eur. Phys. J. C {\bf 80},  145 (2020).

\bibitem{Equilib3} A.B. Balakin and D.E. Groshev, Fingerprints of the Cosmological Constant: Folds in the Profiles of the Axionic Dark Matter Distribution in a Dyon Exterior, Symmetry  {\bf 12}(3), 455  (2020).

\bibitem{Equilib4} A.B. Balakin and D.E. Groshev, Nonminimal dyons with regular gravitational, electric and axion fields, Int. J. Mod. Phys. D {\bf 29}, 2050083 (2020).

\bibitem{Equilib5} A.B. Balakin and A.F. Shakirzyanov, Is the axionic Dark Matter an equilibrium System? Universe {\bf 6}(11), 192 (2020).

\bibitem{GRB17} LIGO Scientific Collaboration, Virgo Collaboration, Fermi Gamma-Ray Burst Monitor, INTEGRAL, Gravitational Waves and Gamma-rays from a Binary Neutron Star Merger: GW170817 and GRB 170817A, Astrophys. J. Lett. {\bf 848}, L13 (2017).


\end{thebibliography}
\end{document}